\let\accentvec\vec 
\documentclass[runningheads,envcountsame,a4paper]{llncs} 
\pdfoutput=1
 
\let\vec\accentvec

\usepackage{llncsdoc}

\usepackage[T1]{fontenc}
\usepackage{lmodern}
\usepackage[english]{babel}
\usepackage[T1]{fontenc}
\usepackage{threeparttable}
\usepackage{graphicx}
\usepackage{url}

\usepackage{ifpdf} 
\ifpdf
\usepackage[final,expansion=true,protrusion=true,spacing=true,kerning=true]{microtype}
\fi
\usepackage{stmaryrd}
\usepackage[x11names, rgb]{xcolor}
\usepackage[utf8]{inputenc}
\usepackage{latexsym}
\usepackage{amsfonts,amssymb,amsmath}
\usepackage{multirow}
\usepackage{booktabs}
\usepackage{tikz}
\usetikzlibrary{decorations,arrows,shapes}
\usetikzlibrary{automata}
\usetikzlibrary{trees}
\usetikzlibrary{calc}

\pgfdeclarelayer{background}
\pgfdeclarelayer{foreground}
\pgfsetlayers{background,main,foreground}

\tikzstyle{sensor}=[draw, fill=blue!20, text width=5em, 
    text centered, minimum height=2.5em]
\tikzstyle{ann} = [above, text width=5em]
\tikzstyle{naveqs} = [sensor, text width=6em, fill=red!20, 
    minimum height=3em, rounded corners]
\def\blockdist{4}
\def\edgedist{2.5}

\usepackage{mathpartir}

\usepackage[final]{pdfpages}
\usepackage{upgreek}
\usepackage{longtable}
\usepackage{lscape}

\usepackage{mathptmx}
\usepackage{lmodern}
\usepackage{helvet}
\usepackage{courier}

\usepackage{makeidx}         
\usepackage{graphicx}        
\usepackage{multicol}        
\usepackage[bottom]{footmisc}

\DeclareFontFamily{T1}{lmtt}{} 
\DeclareFontShape{T1}{lmtt}{m}{n}{<-> ec-lmtl10}{} 
\DeclareFontShape{T1}{lmtt}{m}{\itdefault}{<-> ec-lmtlo10}{} 
\DeclareFontShape{T1}{lmtt}{\bfdefault}{n}{<-> ec-lmtk10}{} 
\DeclareFontShape{T1}{lmtt}{\bfdefault}{\itdefault}{<-> ec-lmtko10}{}

\DeclareFontFamily{OT1}{pzc}{}
\DeclareFontShape{OT1}{pzc}{m}{it}{<-> s * [1.10] pzcmi7t}{}
\DeclareMathAlphabet{\mathpzc}{OT1}{pzc}{m}{it}

\usepackage{algorithm}
\usepackage{algorithmic}

\usepackage{algorithmic,eqparbox,array}
\renewcommand\algorithmiccomment[1]{%
  \hfill\#\ \eqparbox{COMMENT}{#1}%
}
\newcommand\LONGCOMMENT[1]{%
  \hfill\#\ \begin{minipage}[t]{0.4\textwidth}\textit{#1}\strut\end{minipage}%
}
\renewcommand{\algorithmicrequire}{\textbf{Input:}}
\renewcommand{\algorithmicensure}{\textbf{Output:}}

\newenvironment{Algorithm}[2][!ht]%
{\begin{figure*}[#1]
\centering
\begin{minipage}{#2}
\begin{algorithm}[H]}%
{\end{algorithm}
\end{minipage}
\end{figure*}}

\newsavebox{\sembox}
\newlength{\semwidth}
\newlength{\boxwidth}

\usepackage[hidelinks]{hyperref}

\usepackage{listings,xcolor}

\usepackage[normalem]{ulem} 
\newcommand\hl{\bgroup\markoverwith
      {\textcolor{yellow}{\rule[-.5ex]{2pt}{2.5ex}}}\ULon}

\lstnewenvironment{avjada}[1][]
{ \lstset{language=Ada}
    \lstset{escapeinside={(*@}{@*)},
        numbers=left,
  numberstyle=\color{gray}\tiny,
  stepnumber=1,
   breaklines=true,
       framesep=5pt,
       basicstyle=\scriptsize\ttfamily,
       showstringspaces=false,
       keywordstyle=\itshape\color{blue},
       stringstyle=\color{red},
           morecomment=[l][{\color[rgb]{0.1, 0.2, 0.8}}]{\#},
               moredelim=[il][{\color[rgb]{0.1, 0.2, 0.8}}]{@},
    commentstyle=\color{black},
    rulecolor=\color{black},
    xleftmargin=0pt,
    xrightmargin=0pt,
    aboveskip=\medskipamount,
    belowskip=\medskipamount,
           backgroundcolor=\color{white}, #1
}}
{}

\definecolor{lightGrey}{rgb}{0.9, 0.9, 0.9}

\usepackage{tikz}
\usetikzlibrary{shapes,arrows,shadows}
\usetikzlibrary{fit}

\makeatletter
\tikzset{
  fitting node/.style={
    inner sep=0pt,
    fill=none,
    draw=none,
    reset transform,
    fit={(\pgf@pathminx,\pgf@pathminy) (\pgf@pathmaxx,\pgf@pathmaxy)}
  },
  reset transform/.code={\pgftransformreset}
}
\makeatother

\usepackage{tikzscale}
\pgfdeclarelayer{background,foreground}
\pgfsetlayers{background,main,foreground}

\DeclareMathAlphabet{\mathcal}{OMS}{cmsy}{m}{n}

\newenvironment{keywords}{
       \list{}{\advance\topsep by0.35cm\relax\small
       \leftmargin=1cm
       \labelwidth=0.35cm
       \listparindent=0.35cm
       \itemindent\listparindent
       \rightmargin\leftmargin}\item[\hskip\labelsep
                                     \bfseries Keywords:]}
     {\endlist}

 \usepackage{paralist}

\newif\ifcompress
\compresstrue

\ifcompress
\usepackage[small,bf]{caption}

\renewcommand{\figurename}{Fig.}
\renewcommand{\tablename}{Tab.}

\else
\usepackage{caption}
\fi

\usepackage{subcaption}

\usepackage[misc,geometry]{ifsym}

\newif\ifappendix

\usepackage{upquote,textcomp}

\appendixfalse

\begin{document}

\title{Addressing the Regression Test Problem with Change Impact Analysis for
Ada\thanks{The final publication is available at Springer via
\url{http://dx.doi.org/10.1007/978-3-319-39083-3_5}}}

\toctitle{Addressing the Regression Test Problem with Change Impact Analysis for Ada}

\author{Andrew V.\ Jones}
\tocauthor{Andrew V.\ Jones}
\authorrunning{A.\ V.\ Jones}

\institute{Vector Software, Inc.\\London, UK\\
\email{andrew.jones@vectorcast.com}}

\maketitle
\hypersetup{pageanchor=false}

\begin{abstract}
    The \emph{regression test selection problem}---selecting a subset of a
    test-suite given a change---has been studied widely over the past two
    decades.
    However, the problem has seen little attention when constrained to
    high-criticality developments and where a ``safe'' selection of tests need
    to be chosen.
    Further, no practical approaches have been presented for the programming
    language Ada.
    In this paper, we introduce an approach to solving the selection problem
    given a combination of both static and dynamic data for a program and a
    change-set.
    We present a \emph{change impact analysis} for Ada that selects the safe
    set of tests that need to be re-executed to ensure no regressions.
    We have implemented the approach in the commercial, unit-testing tool
    VectorCAST, and validated it on a number of open-source examples.
    On an example of a fully-functioning Ada implementation of a DNS server
    (\textsc{Ironsides}), the experimental results show a 97\% reduction in
    test-case execution.
\end{abstract}

%

\section{Introduction}

In their seminal work of 1988~\cite{HarroldS88}, Harrold \& Soffa introduced a
dataflow-based approach for minimising the regression test effort in the
context of Pascal.
Since then, the problem of regression test execution has seen considerable
attention~\cite{Engstrom+10,Li+13,SoetensDZP15}.

Furthermore, and given the recent emergence of agile
processes~\cite{StalhanHMH14}, which promote test-driven development as well as
continuous integration~\cite{KaistiRMHKML13}, there is now a desire from
developers to be able to re-test modified software rapidly.
However, in the context of Ada, there are few articles (to the best of our
knowledge, there only exists one paper~\cite{Loyall+97} from 1997 that
investigates change impact analysis for Ada) discussing how to solve the
problem, without reverting to ``retest all''~\cite{ParsaiSMD14}.

Consequently, this paper considers the \emph{test-case selection
problem}~\cite{Engstrom+10}:
\begin{quote}
    ``\emph{determine which test-cases need to be re-executed [\ldots\hspace{-0.1em}] in order
    to verify the behaviour of modified software}''
\end{quote}
when applied to systems developed using Ada.
It follows that we aim to investigate the plausibility of applying change
impact analysis to regression testing of Ada source code.
To this end, we seek to minimise the number of tests a developer needs to
re-execute to determine if the behaviour of their software has been affected
after making a change.
%

%

Our approach for \emph{change-based testing} (CBT) of Ada is as follows.
%
We begin by assuming the existence of a test baseline $\mathcal{T}$ of
regression tests associated with a set of Ada source files, as well as access
to both the original and modified source code.
The analysis then proceeds as follows:
\begin{enumerate}
    \item The difference between the original and modified source code is
        assessed to construct a \emph{change-set} $\mathcal{A}$.
        This change-set encapsulates changes at the interface, package and
        subprogram\footnote{In this paper, we use the term ``subprogram'',
        without introducing ambiguity, to refer to either a function or a
    procedure inside of an Ada package.} levels.

    \item An intermediate representation of the program is constructed, based
        on both static data (derived without executing the program) and dynamic
        data (collected by executing the existing test baseline $\mathcal{T}$).
        This intermediate representation forms the basis of a dependency graph
        of the Ada source code.

    \item Given the change-set $\mathcal{A}$ and the intermediate
        representation, we determine a set of tests $\mathcal{T'} \subseteq
        \mathcal{T}$ that is affected by the changes in $\mathcal{A}$. We use
        the internals of the test automation tool VectorCAST to calculate
        the correspondence between changes in $\mathcal{A}$ and the dependency
        graph.

\end{enumerate}
In Step 1, we are concerned with the calculation of the subset of packages and
subprograms that were modified by a given change-set.
Step 2 is focused on establishing the set of interdependencies in the software.
Finally, Step 3 is concerned with the identification of those tests whose
behaviour was affected from the data in Step 1.
As we demonstrate later, we consider the locality (i.e., specification vs. body
vs. subprogram) of the change to allow us to accurately understand its
change-impact.

To-date, approaches to performing a change impact analysis for object-oriented
languages either consider a static or a dynamic-derived dependency
graph~\cite{Engstrom+10,Li+13,SoetensDZP15}.
Uniquely, we consider a hybrid approach, using data from both static and
dynamic analyses.
Our change impact analysis calculates three types of dependency:
\begin{itemize}
    \item Statically:
        \begin{enumerate}
            \item Type and Ada specification dependencies -- where Package
                \texttt{A} depends on Package \texttt{B} as part of
                \texttt{A}'s specification
%
            \item Uses and Ada body dependencies -- where Package \texttt{A}
                depends on Package \texttt{B} as part of \texttt{A}'s body
%
        \end{enumerate}
    \item Dynamically:
        \begin{enumerate}
            \setcounter{enumi}{2}
        \item Subprogram invocation and coupling -- where a subprogram
            \texttt{Foo} in Package \texttt{A} calls a subprogram \texttt{Bar}
            in Package \texttt{B}
        \end{enumerate}
\end{itemize}
Considering dependency data that is derived both statically and dynamically
results in a technique that is not exclusively tied to subprogram-level
analysis~\cite{Orso+04}.
That is, we can consider the change impact at different levels of the software
architecture.
For example, it can support changes that occur at package-scope or to the
object hierarchy.

Approaches based on static slicing~\cite{LarsenH96} of the program are often
overly-conservative, while maintaining ``safety''~\cite{LawR03}.
When developing safety-critical systems, it can be accepted that this
conservatism is of benefit, as it accounts for all possible behaviours of the
system.
However, this can lead to a change impact analysis that results in the
(undesirable) ``retest all'' answer, which can be of little use to developers
wishing to verify their day-to-day work.

Conversely, dynamic slicing (e.g., an analysis based on collected code
coverage), considers only the behaviours and impacts that have been observed as
part of previous system executions.
An analysis based purely on dynamic data will potentially lead to ``unsafe''
conclusions~\cite{LawR03}.

We describe our approach as \emph{safe} -- by this, we mean that any test
contained within ``impact set'' is \emph{at least necessary} to exercise all of
the impacts of the changes in a given change-set.
Our work also aims for \emph{minimality}, but not the \emph{minimal} test-case
set.
Minimality cannot be achieved without a heavier approach to the change-impact
process.
For example, a finer-grained analysis could be based on modifications to the
\emph{def-use chains}~\cite{HummelHN94} for package-level variables, and
subsequently only execute those tests that depend on those variables.

We note that, basing the analysis (partly) on code coverage allows us to avoid
complications when it comes to Ada 83 features such as generics, or Ada 95
features such as dynamic binding~\cite{AdaIC98}.
If the internals of a subprogram change invoke another (late-bound) subprogram,
this would be detected as a subprogram-level change.
Consequently, all tests executing that subprogram would be re-executed,
invoking the newly added dynamic call.
As such, there is no need to adopt a heavier approach that needs to consider
polymorphism~\cite{ParsaiSMD14}.
We discuss this further in Section~\ref{sec:polymorphism}.

\subsubsection*{Structure of the paper.}

The rest of the paper is structured as follows.
In the immediate subsection (Section~\ref{sec:related_work}), we provide an
overview of the relevant literature to the regression test problem.
The subsequent section (\S\ref{sec:background}) provides a brief introduction
to software change impact analysis and VectorCAST.
In Section~\ref{sec:cbt_for_ada}, we introduce our approach to impact analysis
for Ada.
We then provide an experimental evaluation (Section~\ref{sec:eval}), based on a
selection of open-source examples.
In the final section (\S\ref{sec:conclusion}), we conclude.

\subsection{Related Work}\label{sec:related_work}

In 1988, Harrold \& Saffa~\cite{HarroldS88} introduced an \emph{incremental
testing} methodology for Pascal.
To achieve this, they associated a test with the path taken through a module.
The ``incremental tester'' would then try to re-use test-cases by identifying
the tests that exercise the changes, or those which had their execution path
modified by the change. 

Loyall \emph{et al.}~\cite{Loyall+97}, implemented a prototype impact analyser
that presents the static dependency graph in a hyperlinked form to allow for
easy navigation.
While their tool does support Ada, it does not actually calculate the impact of
a change in the source code -- it is designed to support a ``what if'' approach
to potential changes.
A user can select an entity that might be modified, and then see the effects of
this modification.

In~\cite{RenRST05}, Ren \emph{et al.} introduce the tool $\mathit{Chianti}$,
which is able to calculate the set of affecting changes in a Java program that
can lead to the behaviour of a test being modified.
They consider two approaches: one based on static call graphs, and one based on
dynamic call graphs.
However, they do not consider the combination of static and dynamic data for a
more precise analysis.

The theoretical underpinnings of $\mathit{Chianti}$ were presented
in~\cite{RyderT01}, where the classification of \emph{types} of (atomic)
changes in Java programs was introduced.
An approach was then designed to calculate the impact on other areas of the
system, given a collection of atomic changes.

Law \emph{et al.}~\cite{LawR03} consider the application of dynamic program
slicing to the change impact process.
Their approach is focused on the affect of program modifications on other parts
of the program, rather than the test-case minimisation problem.
They present the algorithm $\mathit{PathImpact}$ that decides if a change in
procedure $p$ of a program $P$ has a potential impact on other procedures
reachable from $p$ in the call graph $G$ of $P$.
$\mathit{PathImpact}$ then calculates a forward and backwards slice through the
program, as well as tracking function calls and returns, such that a backwards
analysis is accurately scoped.
In~\cite{OrsoAH03}, Orso \emph{et al.} present the
$\mathit{CoverageImpact}$ algorithm, which walks the execution data in
combination with a forward slice of the variables in the program to calculate
the impacted set.
This set is then used to identify the tests that should be re-executed.

\section{Background}\label{sec:background}

We briefly introduce \emph{change impact analysis} (Section~\ref{sec:cia}) and
VectorCAST (Section~\ref{sec:vectorcast}).

\subsection{Software Change Impact Analysis}\label{sec:cia}

Simply put, software change impact analysis~\cite{Ren+04b} is a family of
techniques for determining the effects and outcomes of a source code
modification, and for improving developer productivity in the context of such a
change.
We refer the interested reader to~\cite{Engstrom+10,Li+13}.

We illustrate the outcome of a potential change in
Figure~\ref{fig:source_tree}.
For example, consider a change to Package~\texttt{C} in the source tree shown.
We will have two types of impact:
\begin{description}
    \item[Upstream changes] -- this is where Package~\texttt{A} calls into
        Package~\texttt{C}. A modification to either the internal behaviour or
        external interface to Package~\texttt{C} can cause a potential change
        in Package~\texttt{A}.
        \item[Downstream changes] -- this is where Package~\texttt{C} calls
            into Package~\texttt{F}. While the internal behaviour of
            Package~\texttt{F} cannot be affected by this change
            (Package~\texttt{F} can be oblivious to Package~\texttt{C}),
            Package~\texttt{F} may now be used in a different way.
\end{description}
In the context of this paper, we are interested in identifying the set of tests
that must be re-run in the presence of a change to Package~\texttt{C}.
To elucidate, any tests that execute directly on \texttt{C} would have to be
re-run (depending on the scope of the change) and any tests associated with
units (e.g., \texttt{A}) that have code coverage on the modified parts of
\texttt{C} should also be re-run.
We exclude re-executing the tests for Package~\texttt{F}, as the tests on
Package~\texttt{C}, which collect coverage on \texttt{F} already, will validate
this modified use of \texttt{F}. 

\tikzstyle{abstract}=[rectangle, draw=black, rounded corners, fill=blue!40, drop shadow,
        text centered, anchor=north, text=white, text width=3cm]
\tikzstyle{comment}=[rectangle, draw=black, rounded corners, fill=green, drop shadow,
        text centered, anchor=north, text=white, text width=3cm]
\tikzstyle{myarrow}=[->, >=open triangle 90, thick]
\tikzstyle{line}=[-, thick]

\begin{center}
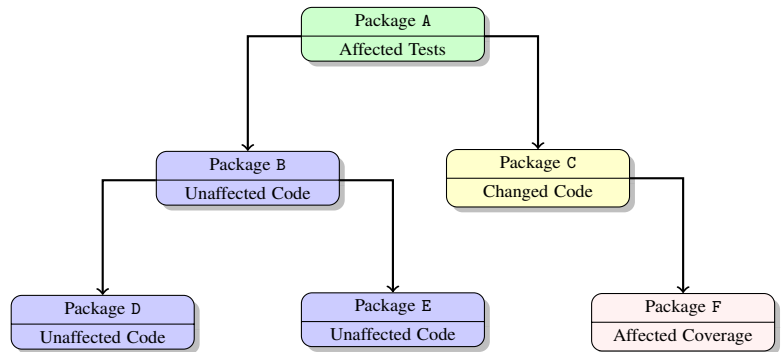
\begin{figure}
\centering
\begin{tikzpicture}[node distance=3.6cm,scale=0.75, transform shape]
    \node (Affected2) [abstract, rectangle split, rectangle split parts=2, fill=green!20,text=black]
        {
            Package~\texttt{A}
		\nodepart{second}Affected Tests
        };
    \node (Changed) [abstract, rectangle split, rectangle split parts=2,fill=yellow!20,text=black,below right of=Affected2]
        {
            Package~\texttt{C}
	    \nodepart{second}Changed Code
        };
    \node (NotAffected) [abstract, rectangle split, rectangle split parts=2, below left of=Affected2,fill=blue!20,text=black]
        {
            Package~\texttt{B}
		\nodepart{second}Unaffected Code
        };
    \node (Affected1) [abstract, rectangle split, rectangle split parts=2, below right of=Changed,fill=pink!20,text=black]
        {
            Package~\texttt{F}
		\nodepart{second}Affected Coverage
        };
    \node (NotAffected3) [abstract, rectangle split, rectangle split parts=2, below left of=NotAffected,fill=blue!20,text=black]
        {
            Package~\texttt{D}
		\nodepart{second}Unaffected Code
        };
    \path (Affected1) -- (NotAffected3) node[midway,yshift=15] (NotAffected2) [abstract, rectangle split, rectangle split parts=2,fill=blue!20,text=black]
        {
            Package~\texttt{E}
		\nodepart{second}Unaffected Code
        };

    \path[->,draw,thick] (Affected2.west)+(0.0,-0.035) -| (NotAffected.north);
    \path[->,draw,thick] (Affected2.east)+(0.0,-0.035) -| (Changed.north);
    \path[->,draw,thick] (Changed.east) -| (Affected1.north);
    \path[->,draw,thick] (NotAffected.west)+(0.0,-0.035) -| (NotAffected3.north);
    \path[->,draw,thick] (NotAffected.east)+(0.0,-0.035) -| (NotAffected2.north);
\end{tikzpicture}
\caption{How changes can propagate through the source tree}\label{fig:source_tree}
\end{figure}
\end{center}

\subsection{VectorCAST}\label{sec:vectorcast}

VectorCAST/Ada\footnote{\url{www.vectorcast.com}; in what follows, we write
VectorCAST to mean VectorCAST/Ada} is a commercial, dynamic unit testing and
code coverage tool for Ada.
To construct automatically unit testing environments for Ada source code,
VectorCAST parses the provided Ada program, extracts the relevant Ada
types/packages, and then presents a ``test-case designer'' that allows a user
to specify tests without the need to write tests in Ada directly.
Crucially, VectorCAST is also able to instrument the source code to obtain code
coverage from test case execution.

%

Following~\cite{PezzeY07}, we note that unit testing environments can be
constructed in two ways:
\begin{itemize}
    \item A ``unit test'' mode, where testing is performed on an individual
        unit, where all of its external dependants have been automatically
        mocked~\cite{PezzeY07}.
    \item An ``integration test'' mode, where testing can be performed across
        multiple units, and where the external dependants have been brought
        into VectorCAST and can be instrumented for code coverage.
        In this mode, the behaviour of the external interfaces (via expected
        call and return values) can also be tested. 
\end{itemize}
With the exception of a change to a dependant specification, change-based
testing in unit testing mode is limited to selecting the tests to re-run inside
of a single unit.
Change impact analysis is more complex when you consider integration-style
tests, as there will be dependencies between the units contained inside the
testing project.
The test selection problem is then to minimise the re-test effort, in the
context of changes in any dependants.

\section{Change-based Testing for Ada}\label{sec:cbt_for_ada}

We now present our approach for performing impact analysis and solving the
test-case selection problem for Ada.

We consider a ``safe'' approach to change impact analysis at the expense of
false negatives: in the context of a safety-critical software development, we
consider it more appropriate to have an overzealous change impact, rather than
exclude a test erroneously (false positives).

\subsection{Dynamic Impact Analysis}

The high-level of a typical \emph{dynamic-only} impact analysis~\cite{Li+13} is
shown in Figure~\ref{fig:dynamic}.
In this figure, we see that the ``core'' of a dynamic impact analysis approach
is the ability to map test data to run-time data, therefore allowing us to
calculate those tests effected.
To support processing the change set into an impact set, we assumed that the
relationship between this data is stored internally in the tool: the
\emph{intermediate representation}.

\pgfdeclarelayer{background}
\pgfdeclarelayer{foreground}
\pgfsetlayers{background,main,foreground}


\tikzstyle{sensor}=[draw, fill=blue!20, text width=5em, 
    text centered, minimum height=2.5em,drop shadow]
\tikzstyle{ann} = [above, text width=5em, text centered]
\tikzstyle{wa} = [sensor, text width=10em, fill=red!20, 
    minimum height=6em, rounded corners, drop shadow]
\tikzstyle{sc} = [sensor, text width=13em, fill=red!20, 
    minimum height=10em, rounded corners, drop shadow]

\def\blockdist{2.3}
\def\edgedist{2.5}

\begin{center}
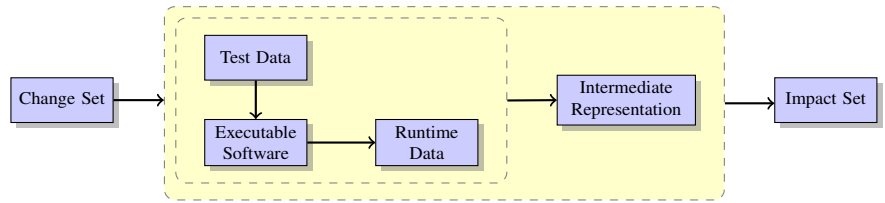
\begin{figure}
\centering

\begin{tikzpicture}[scale=0.75, transform shape]
    
    \node (wa) {};
    
    \path (wa.west)+(-3.0,1.5) node (testdata) [sensor] {Test Data};
    
    \path (wa.west)+(-3.0,0.0) node (exec) [sensor] {Executable Software};
    
    \path (wa.west)+(-0.0,0.0) node (runtime) [sensor] {Runtime Data};
    
    \path [draw, thick, ->] (testdata.south) -- node [above] {} 
    (exec.north);
    
    \path [draw, thick, ->] (exec.east) -- node [above] {} 
    (runtime.west);
        
    \path (testdata.north west)+(-0.5,0.3) node (a) {};
    \path (runtime.south east)+(0.5,-0.3) node (d) {};
    
    \path (wa.west)+(3.5,0.75) node (inter) [sensor, text width=7em] {Intermediate Representation};
    
    \path (wa.west)+(7.0,0.75) node (impact) [sensor] {Impact Set};
    
    \path (wa.west)+(-6.4,0.75) node (change) [sensor] {Change Set};
    
    \path (testdata.north west)+(-0.7,0.5) node (e) {};
    
    \path (d)+(0.5,-0.3) node (d2) {};
    
    \path (inter.south east)+(0.5,-0.3) |- (d2) node[midway] (f) {};
	
    \begin{pgfonlayer}{background}
	\draw[fill=yellow!20,rounded corners, draw=black!50, dashed]
            (e) rectangle (f) node[fitting node] (rect1) {};
	\draw[fill=yellow!20,rounded corners, draw=black!50, dashed]
            (a) rectangle (d) node[fitting node] (rect2) {};
    \end{pgfonlayer}

    \draw[->,thick] (rect2.east) -- (inter.west);
    \draw[->,thick] (change.east) -- (change.east-|rect1.west);
    \draw[->,thick] (rect1.east) -- (rect1.east-|impact.west);
    
\end{tikzpicture}

\caption{Strictly dynamic change impact analysis}\label{fig:dynamic}
\end{figure}

\end{center}

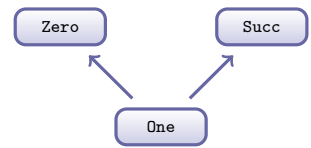
\begin{figure}[ht] 
\captionsetup[subfigure]{position=b}
\vspace{-1em}
\begin{center}
  \begin{subfigure}{0.52\linewidth}
    \begin{minipage}{0.90\textwidth}
\begin{lstlisting}[language=ada]
package body Peano is

   function One return Integer is
   begin
       return Succ(Zero);
   end One;

   function Zero return Integer is
   begin
      return 0;
   end Zero;

   function Succ (Val : in Integer)
       return Integer is
   begin
       return Val + 1;
   end Succ;

end Peano;

\end{lstlisting}
    \end{minipage}
    \vspace*{\fill}
    \caption{A trivial Ada program} 
    \label{fig:peano_adb} 
    \vspace{-1em}
  \end{subfigure}
  \begin{subfigure}{0.32\linewidth}
\begin{tikzpicture}[
    scale=0.75, transform shape,
    node distance=2.5cm,
    grow=down,
    edge from parent/.style={->,very thick,draw=blue!40!black!60,
        shorten >=5pt, shorten <=5pt},
    edge from parent path={(\tikzparentnode.east) -- (\tikzchildnode.west)},
    kant/.style={text width=2cm, text centered, sloped},
    every node/.style={inner sep=2mm},
    punkt/.style={rectangle, rounded corners, shade, top color=white,
    bottom color=blue!50!black!20, draw=blue!40!black!60, very
    thick, minimum width=5em, minimum height=2em }
    ]

\node[punkt] at (0,0) (One) {\texttt{One}};
\node[punkt] (Succ) [above right of=One] {\texttt{Succ}};
\node[punkt] (Zero) [above left of=One] {\texttt{Zero}};

\path[->,very thick,draw=blue!40!black!60,shorten >=5pt, shorten <=5pt] (One) -- (Succ);
\path[->,very thick,draw=blue!40!black!60,shorten >=5pt, shorten <=5pt] (One) -- (Zero);
\end{tikzpicture}
    \caption{Dynamic dependencies}
    \label{fig:peano_deps} 
  \end{subfigure}
\end{center}
  \caption{Example dependencies}
  \label{fig:example_deps} 
\end{figure}

The intermediate representation can take a number of forms when considering a
dynamic analysis.
When considering code coverage-based analyses with information derived from
test execution, such information can be stored as a dynamic dependency tree.
For the Ada program shown in Figure~\ref{fig:peano_adb}, we exemplify its
dynamic-only dependency tree in Figure~\ref{fig:peano_deps}.
A change in either \texttt{Zero} or \texttt{Succ} may affect the behaviour of
\texttt{One}.
    
    \lstset{%
        frame=tb
    }

    \lstset{language=Ada}
    \lstset{escapeinside={(*@}{@*)},
        numbers=left,
  numberstyle=\color{gray}\tiny,
  stepnumber=1,
   breaklines=true,
       framesep=5pt,
       basicstyle=\scriptsize\ttfamily,
       showstringspaces=false,
       keywordstyle=\itshape\color{blue},
       stringstyle=\color{red},
           morecomment=[l][{\color[rgb]{0.1, 0.2, 0.8}}]{\#},
               moredelim=[il][{\color[rgb]{0.1, 0.2, 0.8}}]{@},
    commentstyle=\color{black},
    rulecolor=\color{black},
    xleftmargin=0pt,
    xrightmargin=0pt,
    aboveskip=\medskipamount,
    belowskip=\medskipamount,
           backgroundcolor=\color{white}}

We presume the existence of an original program $P$ and a modified program
$P'$, which has been derived from $P$.
Furthermore, it is also assumed that both $P$ and $P'$ are both syntactically and
semantically correct (i.e., compilable).
The analysis places no restriction beyond these on the nature of the changes.

In the context of what follows, we assume that the intermediate representation
contains both static and dynamic data, and the availability of information
about the packages (specifications and bodies) and subprograms that have been
altered.

\subsection{Intermediate Representation for Ada}

We now introduce the data structures used to construct our analysis for Ada.
As we are developing a hybrid approach using both static and dynamic data, we
introduce both separately.

\subsubsection{Static Data.}

For the data we wish to extract statically from the Ada program, we consider
the following data-types:
\begin{align*}
    \mathit{Contains} &: \mathit{Package} \rightarrow
    {\mathit{Subprogram}^{\ast}} \\
    \mathit{Uses} &: \mathit{Package} \times \left\{ \mathit{Body},
\mathit{Spec} \right\} \rightarrow {\mathit{Package}^{\ast}}
\end{align*}

The data structure $\mathit{Contains}$ is used to map Ada $\mathit{Package}$s
to zero-or-more \textit{Subprogram}s contained within that $\mathit{Package}$.
Similarly, $\mathit{Uses}$ creates a dependency map between $\mathit{Package}$
body and specifications, to the package specifications that they
``\texttt{with}''.

We use the relation $\mathit{Contains}$ to find all affected subprograms given
either a specification or a package body-level change;
$\mathit{Uses}$ allows us to track when a dependant has been modified
(e.g., if package \texttt{A} withs \texttt{B}, and if \texttt{B} changes, we
know that we need to re-execute any test covering package \texttt{A}).

For the presentation that follows, we assume that it is possible to compute the
inverse of $\mathit{Contains}$ and $\mathit{Uses}$.

\subsubsection{Dynamic Data.}

We now consider the dynamic data we require for our analysis:
\begin{align*}
    \mathit{Covers}: \mathit{Test} \rightarrow {\mathit{Subprogram}}^{\ast}
\end{align*}
which maps test-cases in the test baseline, $\mathcal{T}$, to the subprograms
covered when a given test is executed.
We note that, unlike~\cite{LawR03,Orso+04}, we are not concerned with the
ordering of subprogram calls/returns for a given test.

It is clear that, when combining these tree-like data structures, it is
possible to construct a combined, static/dynamic dependency tree.
Such a tree could be unfolded to construct a directed, acyclic dependency graph
of the program.
This is because dependency relationships between entities are transitive.
That is, if \texttt{A} depends on \texttt{B} and \texttt{B} depends on
\texttt{C} in one or more dependency relationships, then \texttt{A} depends on
\texttt{C}.

\subsection{Example}

Before presenting the approach to solve the test-case selection problem, we
exemplify the technique when applied to Ada source code.
We illustrate the process using the small Ada program shown in
Figure~\ref{fig:ada_example}.

In this example, we have two packages (\texttt{A} and \texttt{B}), each
containing a single function.
In the body of package \texttt{A}, we have an external dependency on the
specification of \texttt{B}, via the use of the ``\texttt{with}'' directive.
It is clear that there is an implicit dependency between each package and its
specification (i.e., that the body of \texttt{A} depends on the specification
of \texttt{A}).
It follows that we have $\texttt{A} \times \mathit{Body} \rightarrow
\texttt{B}$ in $\mathit{Uses}$, and $\texttt{A} \rightarrow \texttt{Foo}$ in
$\mathit{Contains}$.

    \lstset{%
        frame=tb
    }

    \lstset{language=Ada}
    \lstset{escapeinside={(*@}{@*)},
        numbers=left,
  numberstyle=\color{gray}\tiny,
  stepnumber=1,
   breaklines=true,
       framesep=5pt,
       basicstyle=\scriptsize\ttfamily,
       showstringspaces=false,
       keywordstyle=\itshape\color{blue},
       stringstyle=\color{red},
           morecomment=[l][{\color[rgb]{0.1, 0.2, 0.8}}]{\#},
               moredelim=[il][{\color[rgb]{0.1, 0.2, 0.8}}]{@},
    commentstyle=\color{black},
    rulecolor=\color{black},
    xleftmargin=0pt,
    xrightmargin=0pt,
    aboveskip=\medskipamount,
    belowskip=\medskipamount,
           backgroundcolor=\color{white}}

\begin{figure}[ht] 
\captionsetup[subfigure]{position=b}
\vspace{-1em}
  \begin{subfigure}{0.5\linewidth}
    \centering
    \begin{minipage}{0.7\textwidth}
\begin{lstlisting}[language=ada]
package A is

    function Foo
        return Integer; 

end A;
\end{lstlisting}
    \end{minipage}
    \vspace*{\fill}
    \caption{Package Specification for \texttt{A}} 
    \label{fig:package_spec_a} 
    \vspace{4ex}
  \end{subfigure}
  \begin{subfigure}{0.5\linewidth}
    \centering
    \begin{minipage}{0.9\textwidth}
\begin{lstlisting}[language=ada]
with B;

package body A is

    Qux : Integer;

    function Foo return Integer is
    begin
        return Qux + B.Bar;
    end;

begin

    Qux := 0;

end A;
\end{lstlisting}
    \end{minipage}
    \caption{Package Body for \texttt{A}} 
    \label{fig:package_body_a} 
    \vspace{4ex}
  \end{subfigure} 
  \begin{subfigure}{0.5\linewidth}
    \centering
    \begin{minipage}{0.7\textwidth}
\begin{lstlisting}[language=ada]
package B is

    function Bar
        return Integer; 

end B;
\end{lstlisting}
    \end{minipage}
    \vspace*{\fill}
    \caption{Package Specification for \texttt{B}} 
    \label{fig:package_spec_b} 
  \end{subfigure}
  \begin{subfigure}{0.5\linewidth}
    \centering
    \begin{minipage}{0.9\textwidth}
\begin{lstlisting}[language=ada]
package body B is

    Narf : Integer;

    function Bar return Integer is
    begin
        return Narf;
    end;

begin

    Narf := 0;

end B;
\end{lstlisting}
    \end{minipage}
    \caption{Package Body for \texttt{B}} 
    \label{fig:package_body_b} 
  \end{subfigure} 
  \caption{An exemplary Ada program}
  \label{fig:ada_example} 
\vspace{-1em}
\end{figure}

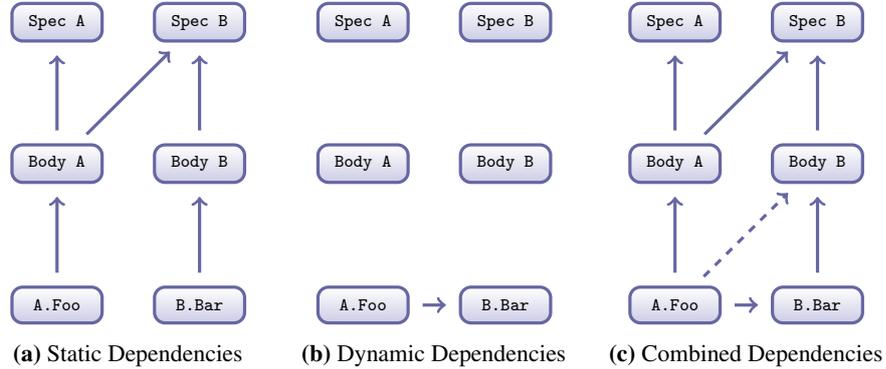
\begin{figure}[ht] 
\captionsetup[subfigure]{position=b}
  \begin{subfigure}{0.33\linewidth}
    \centering
\begin{tikzpicture}[
    scale=0.75, transform shape,
    node distance=2.5cm,
    grow=down,
    edge from parent/.style={->,very thick,draw=blue!40!black!60,
        shorten >=5pt, shorten <=5pt},
    edge from parent path={(\tikzparentnode.east) -- (\tikzchildnode.west)},
    kant/.style={text width=2cm, text centered, sloped},
    every node/.style={inner sep=2mm},
    punkt/.style={rectangle, rounded corners, shade, top color=white,
    bottom color=blue!50!black!20, draw=blue!40!black!60, very
    thick, minimum width=5em, minimum height=2em }
    ]

\node[punkt] at (0,0) (SpA) {\texttt{Spec A}};
\node[punkt] (SpB) [right of=SpA] {\texttt{Spec B}};
\node[punkt] (BdA) [below of=SpA] {\texttt{Body A}};
\node[punkt] (BdB) [below of=SpB] {\texttt{Body B}};
\node[punkt] (Foo) [below of=BdA] {\texttt{A.Foo}};
\node[punkt] (Bar) [below of=BdB] {\texttt{B.Bar}};
\path[->,very thick,draw=blue!40!black!60,shorten >=5pt, shorten <=5pt] (BdA) -- (SpA);
\path[->,very thick,draw=blue!40!black!60,shorten >=5pt, shorten <=5pt] (BdA) -- (SpB);
\path[->,very thick,draw=blue!40!black!60,shorten >=5pt, shorten <=5pt] (BdB) -- (SpB);
\path[->,very thick,draw=blue!40!black!60,shorten >=5pt, shorten <=5pt] (Foo) -- (BdA);
\path[->,very thick,draw=blue!40!black!60,shorten >=5pt, shorten <=5pt] (Bar) -- (BdB);
\end{tikzpicture}
    
    \caption{Static Dependencies}
    \label{fig:static_deps} 
  \end{subfigure}
  \begin{subfigure}{0.33\linewidth}
    \centering
\begin{tikzpicture}[
    scale=0.75, transform shape,
    node distance=2.5cm,
    grow=down,
    edge from parent/.style={->,very thick,draw=blue!40!black!60,
        shorten >=5pt, shorten <=5pt},
    edge from parent path={(\tikzparentnode.east) -- (\tikzchildnode.west)},
    kant/.style={text width=2cm, text centered, sloped},
    every node/.style={inner sep=2mm},
    punkt/.style={rectangle, rounded corners, shade, top color=white,
    bottom color=blue!50!black!20, draw=blue!40!black!60, very
    thick, minimum width=5em, minimum height=2em }
    ]

\node[punkt] at (0,0) (SpA) {\texttt{Spec A}};
\node[punkt] (SpB) [right of=SpA] {\texttt{Spec B}};
\node[punkt] (BdA) [below of=SpA] {\texttt{Body A}};
\node[punkt] (BdB) [below of=SpB] {\texttt{Body B}};
\node[punkt] (Foo) [below of=BdA] {\texttt{A.Foo}};
\node[punkt] (Bar) [below of=BdB] {\texttt{B.Bar}};
\path[->,very thick,draw=blue!40!black!60,shorten >=5pt, shorten <=5pt] (Foo) -- (Bar);
\end{tikzpicture}
    \caption{Dynamic Dependencies}
    \label{fig:dynamic_deps} 
  \end{subfigure} 
  \begin{subfigure}{0.33\linewidth}
    \centering
\begin{tikzpicture}[
    scale=0.75, transform shape,
    node distance=2.5cm,
    grow=down,
    edge from parent/.style={->,very thick,draw=blue!40!black!60,
        shorten >=5pt, shorten <=5pt},
    edge from parent path={(\tikzparentnode.east) -- (\tikzchildnode.west)},
    kant/.style={text width=2cm, text centered, sloped},
    every node/.style={inner sep=2mm},
    punkt/.style={rectangle, rounded corners, shade, top color=white,
    bottom color=blue!50!black!20, draw=blue!40!black!60, very
    thick, minimum width=5em, minimum height=2em }
    ]

\node[punkt] at (0,0) (SpA) {\texttt{Spec A}};
\node[punkt] (SpB) [right of=SpA] {\texttt{Spec B}};
\node[punkt] (BdA) [below of=SpA] {\texttt{Body A}};
\node[punkt] (BdB) [below of=SpB] {\texttt{Body B}};
\node[punkt] (Foo) [below of=BdA] {\texttt{A.Foo}};
\node[punkt] (Bar) [below of=BdB] {\texttt{B.Bar}};
\path[->,very thick,draw=blue!40!black!60,shorten >=5pt, shorten <=5pt] (BdA) -- (SpA);
\path[->,very thick,draw=blue!40!black!60,shorten >=5pt, shorten <=5pt] (BdA) -- (SpB);
\path[->,very thick,draw=blue!40!black!60,shorten >=5pt, shorten <=5pt] (BdB) -- (SpB);
\path[->,very thick,draw=blue!40!black!60,shorten >=5pt, shorten <=5pt] (Foo) -- (BdA);
\path[->,very thick,draw=blue!40!black!60,shorten >=5pt, shorten <=5pt] (Bar) -- (BdB);
\path[->,very thick,draw=blue!40!black!60,shorten >=5pt, shorten <=5pt] (Foo) -- (Bar);
\path[->,very thick,draw=blue!40!black!60,shorten >=5pt, shorten <=5pt, dashed] (Foo) -- (BdB);
\end{tikzpicture}
    
    \caption{Combined Dependencies}
    \label{fig:combined_deps} 
  \end{subfigure}
  \caption{Types of dependency}
  \label{fig:types_of_dep} 
\end{figure}

For the Ada example illustrated in Figure~\ref{fig:ada_example}, we show the
\emph{static-only dependencies} (i.e., those excluding subprogram calls) in
Figure~\ref{fig:static_deps}.
As we can see, when we do not consider subprogram invocations between packages,
there is no statically-determined dependency between \texttt{A}'s package body
and \texttt{B}'s package body.

We now consider that a test-case $t$ has been created that exercises the
subprogram \texttt{Foo}.
%
%
In this instance, dynamically executing a test-case for the function
\texttt{Foo} will then obtain code coverage on both \texttt{Foo} and
\texttt{Bar}.
After $t$ has executed, we can see that there is a (dynamic) dependency between
\texttt{Foo} and \texttt{Bar} (Figure~\ref{fig:dynamic_deps}).
That is, we have $t \rightarrow \left\{\texttt{Foo}, \texttt{Bar} \right\}$ in
$\mathit{Covers}$.

Finally, the combined dependencies are show in Figure~\ref{fig:combined_deps}.
As we can see, this is the union of the dependencies from the static and the
dynamic data.
As shown in Figure~\ref{fig:combined_deps}, there now exists an \emph{implied}
dependency between \texttt{Foo} and the body of \texttt{B} (the dashed arrow
between \texttt{Foo} and \texttt{B}).
This is because we have a traversal through the dependency graph of:
\[
    \texttt{Foo} \rightarrow \texttt{Bar} \rightarrow \texttt{B}
\]
Consequently, it can be calculated\footnote{where ``\emph{impact}'' is the
inverse relation of dependency.} that a change the body of \texttt{B} will
impact test-cases that are associated with the subprogram \texttt{Foo}.

\subsection{Calculating the Selection}\label{sec:computing}

To solve the test-case selection problem, we introduce an ancillary algorithm
\textsc{AffectedSubprograms} (Algorithm~\ref{alg:affected}).
The algorithm is a classic \emph{work-list} algorithm, used to calculate the
transitive closure of the dependency tree.
For ease, we use $\mathit{entity}$ to refer to a specification, body or
subprogram.

\begin{Algorithm}{0.83\textwidth}
    
    \caption{\textsc{AffectedSubprograms}}
    \label{alg:affected}
    \begin{algorithmic}[1]
    \REQUIRE $\mathit{change} : \mathit{entity}$ \LONGCOMMENT{change entity}
    \REQUIRE $\mathit{static\_dependencies} : \mathit{entity} \rightarrow {\mathit{entity}}^{{\ast}}$ \LONGCOMMENT{static dependencies}
    \ENSURE $\mathit{impacted\_subprograms}$ \LONGCOMMENT{set of affected subprograms} 
    \vspace{0.2em}\hrule\vspace{0.2em}

    \STATE $\mathit{impacted\_subprograms} \gets \emptyset$
    \STATE $\mathit{found} \gets \emptyset$
    \STATE $\mathit{new} \gets \left\{ change \right\}$

    \WHILE{$\mathit{new} \neq \emptyset$}

        \STATE $\mathit{next} \gets \mathit{new.pop()}$ \LONGCOMMENT{pops and removes}
        \STATE $\mathit{found} \gets \mathit{found} \cup \left\{ \mathit{next} \right\}$

        \IF{$\mathit{next}$ is $\mathit{subprogram}$}

            \STATE $\mathit{impacted\_subprograms} \gets \mathit{impacted\_subprograms} \cup \left\{ \mathit{next} \right\}$

        \ENDIF

        \STATE $\mathit{successors} \gets \mathit{static\_dependencies}\left( \mathit{next} \right)$
        \STATE $\mathit{unprocessed} \gets \mathit{successors} \setminus \mathit{found}$ 
        \STATE $\mathit{new} \gets \mathit{new} \cup \mathit{unprocessed}$

    \ENDWHILE

    \RETURN $\mathit{impacted\_subprograms}$
    \end{algorithmic}
\end{Algorithm}

Our algorithm for solving the test-case selection problem is shown in
Algorithm~\ref{alg:affected_tests}.
The algorithm takes a given Ada program $P$, a baseline set of tests
$\mathcal{T}$, the data stored in $\mathit{Covers}$ and changed entity $c$, and
returns the set of tests to be re-executed.
Once the set of affected subprograms has been computed by
$\textsc{AffectedSubprograms}$, $\textsc{AffectedTests}$ iterates over these
subprograms and selects all tests covering them.
These selected tests represent our solution to the test-case selection problem.

We note that $\textsc{AffectedTests}$ relies on an external procedure
$\textsc{StaticDep}$, which calculates the transitive closure of
$\mathit{Contains}$ and $\mathit{Uses}$.

\begin{Algorithm}{0.82\textwidth}
    
    \caption{\textsc{AffectedTests}}
    \label{alg:affected_tests}
    \begin{algorithmic}[1]
    \REQUIRE $P$ \LONGCOMMENT{an Ada program}
    \REQUIRE $\mathcal{T}$ \LONGCOMMENT{a set of tests}
    \REQUIRE $\mathit{Covers}: \mathcal{T} \rightarrow {\mathit{Subprograms}}^{\ast}$ \LONGCOMMENT{test coverage}
    \REQUIRE $c: \mathit{entity}$ \LONGCOMMENT{a change in $P$}
    \ENSURE $\mathit{impacted\_tests} \subseteq \mathcal{T}$ \LONGCOMMENT{set of affected tests} 
    \vspace{0.2em}\hrule\vspace{0.2em}

    \STATE $\mathit{impacted\_tests} \gets \emptyset$
    
    \STATE $\mathit{impacted\_subprograms} \gets \textsc{AffectedSubprograms}\left( c, \textsc{StaticDep}\left(P\right)\right)$

    \FORALL{$t \in \mathcal{T}$}

        \FORALL{$m \in \mathit{impacted\_subprograms}$}

            \IF{$m \in \mathit{Covers}\left(t\right)$}

                \STATE $\mathit{impacted\_tests} \gets \mathit{impacted\_tests} \cup \left\{ t \right\}$

                \STATE $\mathbf{break}$

            \ENDIF

        \ENDFOR

    \ENDFOR

    \RETURN $\mathit{impacted\_tests}$
    \end{algorithmic}
\end{Algorithm}

Given a change-set comprising of a number of modifications to the program
(e.g., multiple package body or subprogram changes), it is possible to
encapsulate \textsc{AffectedTests} in a higher-level procedure that iterates
over each change and collects the aggregate set of affected tests (c.f.,
$\mathit{ImpactAnalysis}$ in~\cite{OrsoAH03}).

\subsection{On Change Impact for Polymorphic Programs}\label{sec:polymorphism}

There has been a lot of consideration in
literature~\cite{HuangS07,Ren+04b,RyderT01} applied to the intricacies of
change impact pertaining to object oriented programming.
However, in the context of the framework presented, the use of object oriented
techniques within Ada does not introduce any further difficulties.

For example, consider a change $C$ that affects the dynamic call tree in a
given program $P$.
We will consider the addition or removal of a specialised subprogram in a
derived package.
If a specialised subprogram is added/removed from a derived package, then the
derived specification (upon which $P$ depends) will change, leading to all
tests for $P$, which have code coverage on the derived package, to be
re-executed.

If a package body member is changed in the base package, then this will
invalidate all tests that have associated code coverage on the derived package,
if the derived package has any static/dynamic calls to its parent.
If there are no tests that generate any coverage on the base package via calls
from the derived package, then a modification to the package global in the base
package will have no effect on the derived package's behaviour, and so no tests
will be impacted.

\subsubsection{Example.}

\begin{figure}[ht] 
    \centering
\begin{tikzpicture}[
    scale=0.75, transform shape,
    node distance=2.5cm,
    grow=down,
    edge from parent/.style={->,very thick,draw=blue!40!black!60,
        shorten >=5pt, shorten <=5pt},
    edge from parent path={(\tikzparentnode.east) -- (\tikzchildnode.west)},
    kant/.style={text width=2cm, text centered, sloped},
    every node/.style={inner sep=2mm},
    punkt/.style={rectangle, rounded corners, shade, top color=white,
    bottom color=blue!50!black!20, draw=blue!40!black!60, very
    thick, minimum width=5em, minimum height=2em }
    ]

\node[punkt] at (0,0) (SpB) {\texttt{Spec Base}};
\node[punkt] (BdB) [below of=SpB] {\texttt{Body Base}};
\node[punkt] (SpD) [right of=BdB] {\texttt{Spec Der.}};
\node[punkt] (BdD) [below of=SpD] {\texttt{Body Der.}};
\node[punkt] (BtaB) [below of=BdB] {\texttt{B.Beta}};
\node[punkt] (AlpB) [left of=BtaB] {\texttt{B.Alpha}};
\node[punkt] (PgP) [right of=BdD] {\texttt{Prog.\ P}};
\node[punkt,bottom color=red!50!black!20, draw=red!40!black!60] (test) [right of=PgP] {\texttt{Test t}};
\node[punkt] (AlpD) [below of=BdD] {\texttt{D.Alpha}};
\path[->,very thick,draw=blue!40!black!60,shorten >=5pt, shorten <=5pt] (BdB) -- (SpB);
\path[->,very thick,draw=blue!40!black!60,shorten >=5pt, shorten <=5pt] (SpD) -- (SpB);
\path[->,very thick,draw=blue!40!black!60,shorten >=5pt, shorten <=5pt] (BdD) -- (SpD);
\path[->,very thick,draw=blue!40!black!60,shorten >=5pt, shorten <=5pt] (BtaB) -- (BdB);
\path[->,very thick,draw=blue!40!black!60,shorten >=5pt, shorten <=5pt] (AlpB) -- (BdB);
\path[->,very thick,draw=blue!40!black!60,shorten >=5pt, shorten <=5pt] (AlpD) -- (BdD);
\path[->,very thick,draw=blue!40!black!60,shorten >=5pt, shorten <=5pt, dashed] (AlpD) -- (BtaB);
\path[->,very thick,draw=blue!40!black!60,shorten >=5pt, shorten <=5pt] (PgP) -- (SpD);
\path[->,very thick,draw=blue!40!black!60,shorten >=5pt, shorten <=5pt, dashed] (PgP) -- (AlpD);
\path[->,very thick,draw=blue!40!black!60,shorten >=5pt, shorten <=5pt, dotted] (test) -- (PgP);
\end{tikzpicture}
    
    \caption{A polymorphic dependency tree}
    \label{fig:polymorphism} 
  \end{figure}
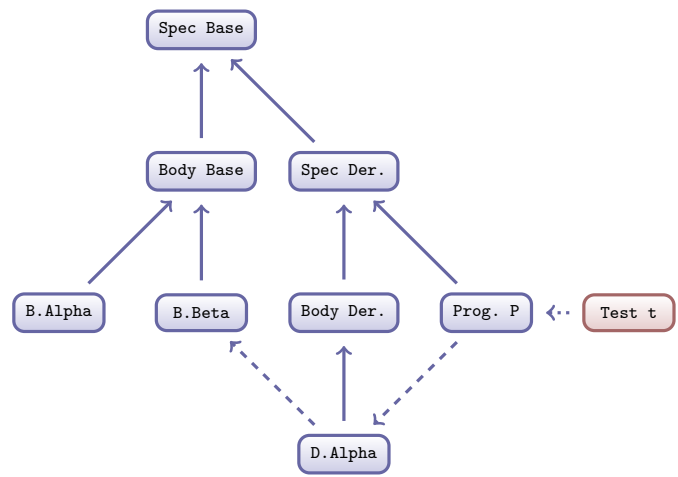

Consider two packages \texttt{Base} and \texttt{Derived}, where the
specification of \texttt{Base} has two subprograms \texttt{Alpha} and
\texttt{Beta}, and that \texttt{Derived} only specialises the subprogram
\texttt{Alpha}.
We further assume a program $P$, and associated test, that calls
\texttt{Derived.Alpha}, and \texttt{Derived.Alpha} calls \texttt{Base.Beta}.
This will create a combined dependency tree as shown in
Figure~\ref{fig:polymorphism} (we use a dashed line to show dynamic
dependencies).

If we now extended \texttt{Derived} such that it contains a specialised version
of \texttt{Beta}, this would then cause a change in the specification and body
of \texttt{Derived}, and so we would re-execute any tests that have coverage on
the subprogram \texttt{Alpha}.

Alternatively, consider a change to a package body member in \texttt{Base}.
Via the dependency tree from Figure~\ref{fig:polymorphism}, this would then
cause any tests with coverage on \texttt{Base.Alpha} and \texttt{Base.Beta} to
be invalidated.
Consequently, our test on \texttt{Derived.Alpha} would therefore be affected,
as per the dynamic coverage collected.

\section{Experimental Evaluation}\label{sec:eval}

To validate the effectiveness of the technique presented in reducing the number
of test-cases to be re-executed, we performed an empirical evaluation comparing
VectorCAST with and without change impact analysis.

\subsection{Experimental Setup}

We considered examples from two sources: ``Malaise'' and \textsc{Ironsides}; we
summarise these below.
A high-level overview of the packages selected is shown in
Table~\ref{tab:specifics}.

{
\centering
\begin{table}
\begin{center}
\small
\caption{Example specifics}\label{tab:specifics}
\begin{tabular}{l @{\hspace{2em}} l @{\hspace{2em}} l}
\toprule
Metric & Malaise & \textsc{Ironsides} \\
\midrule

Number of files & 9 & 9 \\

Number of lines (incl.\ comments/whitespace) & 654 & 4,745 \\

Number of non-empty Ada lines & 468 & 3,441 \\

Number of subprograms & 46 & 97 \\

Aggregate complexity metric~\cite{Watson+96} & 94 & 492 \\

Total number of tests & 228 & 573 \\

Coverage (statement / branch) & 68\% / 68\% & 47\% / 36\% \\

\bottomrule
\end{tabular}
\end{center}
\vspace{-1em}
\end{table}
}

\subsubsection{Malaise.}\label{sec:malaise}
We considered a selection of 9 files taken from~\cite{Malaise15} -- a copy-left
repository of Unix-based utilities written in Ada.
Some of the packages selected included: \texttt{ada\_words.adb}, which provides
``basic Ada parsing of delimiters, separators and reserved words'';
\texttt{conditions.adb} that supports ``several tasks to wait until unblocked
all together or one by one''; and \texttt{forker.adb}, an ``API to a standalone
forker process''.


%
%
%
%
%
%

\subsubsection{\textsc{Ironsides}.}\label{sec:ironsides}

The Internet Domain Name System---or DNS---is an infrastructure whose
responsibility it is to translate domain names (e.g.,
\href{http://www.vectorcast.com}{\texttt{www.vectorcast.com}}) into their
corresponding IP addresses (e.g.,
\href{http://www.vectorcast.com}{\texttt{67.225.168.102}}).
\textsc{Ironsides}~\cite{FaginC13}, an open-source and freely-available DNS
server implemented in SPARK Ada.
Via the use of SPARK, the code of \textsc{Ironsides} is mathematically proven
to be free of defects via the use of formal methods.
For the purposes of this evaluation, we consider a subset of 9 files taken from
the \textsc{Ironsides} ``authoritative'' (2015-04-15) branch~\cite{Carlisle15}.

\subsubsection{Testing methodology.}

To support the empirical evaluation of the presented change-based testing
approach, we used VectorCAST to generate automatically three types of test:
\vspace{-0.5em}
\begin{itemize}
    \item ``empty tests'' -- these are default test-cases generated by
        VectorCAST that provide empty parameter values to every function;
    \item ``min-mid-max tests'' -- these call each test with the min, mid and
        maximum value for each parameter;
    \item ``basis path tests'' -- we used VectorCAST's ability to generate
        automatically basis path tests according to McCabe's complexity
        metric~\cite{Watson+96}.
\end{itemize}
\vspace{-0.5em}
For Malaise, we generated all three types of test; however, to produce a
manageable test-suite size, we only generated empty and basis path tests for
\textsc{Ironsides} (i.e., we did not consider min-mid-max tests).
The size of the test-suite and the coverage attained from its execution are
presented in Table~\ref{tab:specifics}.

For each of the examples, we used VectorCAST to capture the initial state of
the software, and then applied modifications to each of the files: namely, we
added a ``\texttt{null;}'' statement to the beginning of a number of
subprograms, such that VectorCAST would detect a subprogram-level change.
An example of an automated change---highlighted with a box---to the package
\texttt{Ada\_Words} from Malaise is shown in Listing~\ref{fig:ada_words_mod}.

\begin{figure}
\begin{center}
\begin{minipage}{0.72\textwidth}
\lstset{%
    caption=An example modification in the package \texttt{Ada\_Words}\label{fig:ada_words_mod},
    basicstyle=\ttfamily\footnotesize\bfseries,
    frame=tb
}

\begin{avjada}
function Is_Delimiter (C : Character) return Boolean is
begin
   (*@\fbox{\hspace{1ex}null;\hspace{1ex}}@*)
    case C is
        when '&' | '(*@\hspace{0.13ex}{\color{red}{'}}\hspace{0.13ex}@*)' | '(' | ')' | '*' | '+' |
             ',' | '-' | '.' | '/' | ':' | ';' |
             '<' | '=' | '>' | '|' =>
	    return True;
	when others =>
	    return False;
    end case;
end Is_Delimiter;
\end{avjada}
\end{minipage}
\end{center}
\end{figure}

After applying each change, we then performed an ``incremental build and
execute'' inside of VectorCAST, to analyse the code-base and then only re-test
the code that changed.
To validate the effectiveness of the proposed approach, we executed the same
process but without passing the incremental flag to VectorCAST.
The version of VectorCAST used for both the incremental and non-incremental
runs was the official release of 6.4d (released 2016-02-29).

All of the Ada sources for both of the examples (reproduced under a copy-left
licence from both~\cite{Malaise15} and~\cite{Carlisle15}), the VectorCAST
artefacts (e.g., the auto-generated tests) and an ``evaluation runner'' script
are available from~\cite{Jones16}.

\subsection{Results}

We performed our evaluation on a 32-bit Linux machine running Fedora 21, with 8
GiB of RAM and a 6-core Intel Xeon clocked at 2.50GHz.
The compiler used was ``GNAT 4.9.2 20150212 (Red Hat 4.9.2-6)''.

{
    \vspace{-1em}
\centering
\begin{table}
\begin{center}
\small
\caption{Experimental Results}\label{tab:eval}
\begin{tabular}{r @{\hspace{2em}} r @{\hspace{2em}} c @{\hspace{2em}} c @{\hspace{2em}} c @{\hspace{2em}} c }
\toprule
\multirow{2}{*}{Example} & \multirow{2}{*}{Mode} & Units & Subprograms & \# Tests & Build + Exec. \\ 
                         & & Changed & Changed & Executed & Time (s) \\
\midrule
\multirow{2}{*}{Malaise} & Without CBT & \multirow{2}{*}{9} & \multirow{2}{*}{21} & 4,788 & 1,002.48 \\
                         & With CBT & & & 165 & 165.85 \\
\midrule
\multirow{2}{*}{\textsc{Ironsides}} & Without CBT & \multirow{2}{*}{9} & \multirow{2}{*}{93} & 53,289 & 6,986.17 \\
& With CBT & & & 1,347 & 1,147.14 \\
\bottomrule
\end{tabular}
\end{center}
\vspace{-1em}
\end{table}
}

The results of our evaluation can be seen in Table~\ref{tab:eval}.
%
%
The column ``\# Tests Executed'' represents the total number of tests
re-executed after performing the individual subprogram change, with each change
processed separately.
Similarly, ``Build + Exec.\ Time'' is the total time (in seconds) that
VectorCAST took to re-build the test environment, incorporating the current
change-set, and to re-run the affected tests.

As we can see, using the change impact analysis presented in this paper, the
total number of tests needing to be executed for Malaise was reduced from 4,788
(running all 228 test-cases for each of the 21 changes) to only 165 (re-running
only the impacted tests).
Similarly, for \textsc{Ironsides}, the number of tests required to be
re-executed to ensure that no regressions were introduced in the software was
reduced by 97\%.

We observe that the final column (time) does not scale accordingly, as the
auto-generated tests are quick to execute, compared to the higher-cost
environment construction.
Nonetheless, across both examples, we see an 84\% reduction in time to re-test.

Given the size and real-world applicability of \textsc{Ironsides} (with its
higher performance than commercial DNS servers~\cite{FaginC13}),
%
%
we feel that the results obtained would be representative of the benefits
achievable in an industrial Ada project.


\section{Conclusions}\label{sec:conclusion}

In this paper, we have introduced the first practical approach to applying
change impact analysis to the test-case selection problem for Ada.
To the best of our knowledge (c.f.,~\cite{Engstrom+10,Li+13}), ours is the
first approach that explicitly uses a combination of both statically derived
data and dynamic data from test execution.
In safety-critical markets (see, e.g., DO-178C~\cite{DO178C} for aeronautics),
it is commonplace for there to be a requirement to demonstrate ``test
completeness'' via a code coverage mandate.
Consequently, linking a change-impact analysis to data that engineers will
already be collecting is advantageous.

We also considered the affect of object oriented techniques when identifying
those tests to be re-executed.
Considering exclusively static data has previously been
investigated~\cite{RyderT01} and lead to a number of ``heavy-weight''
frameworks~\cite{RenRST05}.
While simplistic, our approach can also handle changes introduced in the
polymorphic hierarchy.

We performed an empirical evaluation of our technique as part of an
experimental extension to VectorCAST.
Our results on a modest-sized example are promising, but further evaluation is
needed.

\subsection{Further Work}

We have identified a number of additional avenues that could improve on the
test-case selection process (at the expense of a heavier technique).
The most immediate area to tackle is on the change impact process at a lower
level than just subprograms.
For example, if the change is constrained to a particular branch of
a conditional, then it would be plausible, without a loss of safety, to select
only those tests that previously entered the same block.

Our presentation of $\mathit{AffectedSubprograms}$, and calculating the
transitive closure of $\mathit{StaticDep}$ (Section~\ref{sec:computing}), leads
us to invalidate all tests when the change is associated to a package
specification or a body.
When we consider a body-level change that, e.g., changes or introduced a new
body-level member variable, this leads us to re-execute more tests than
necessary.
If we considered only those subprograms that referred to each member variable,
we could then be more selective with those that we invalidate.

We leave consideration of how to efficiently handle type modifications at the
specification level for further work.

\subsection{Closing Remarks}

In this paper, we presented, to the best of our knowledge, the first approach
for considering change impact analysis for Ada applied to regression testing
(outside of~\cite{Loyall+97}, which did not consider the test case
selection problem).
As highlighted above, there are a number of improvements to this technique to
further reduce the scope of selected changes.
We position this work as the first footing in this direction, and are not
discouraged by the modest framework presented.

\bibliographystyle{splncs03}
\bibliography{ada_europe_2016}

\end{document}